\documentclass{PoS}

\usepackage{xspace}
\newcommand{\Bd}{\ensuremath{B^0}\xspace}
\newcommand{\Bu}{\ensuremath{B^+}\xspace}
\newcommand{\Bs}{\ensuremath{B_s^0}\xspace}
\newcommand{\B}{\ensuremath{B_{(s)}^0}\xspace}
\newcommand{\Bdmm}{\ensuremath{\Bd \rightarrow \mu^+\mu^-}\xspace}
\newcommand{\Bsmm}{\ensuremath{\Bs \rightarrow \mu^+\mu^-}\xspace}
\newcommand{\Bmm}{\ensuremath{\B \rightarrow \mu^+\mu^-}\xspace}
\newcommand{\BuJpsiK}{\ensuremath{\Bu \rightarrow J/\psi K^+}\xspace}
\newcommand{\Br}{\ensuremath{\mathcal{B}}\xspace}

\newcommand{\Mmm}{\ensuremath{M_{\mu^+\mu^-}}\xspace}
\newcommand{\Bhh}{\ensuremath{\B \rightarrow h^+h^{\prime-}}\xspace}

\title{Updated Search for \Bmm at CDF}

\ShortTitle{Updated Search for \Bmm at CDF}

\author{\speaker{Thomas Kuhr} for the CDF collaboration\\
        Institut f\"ur Experimentelle Kernphysik, KIT\\
        E-mail: \email{Thomas.Kuhr@kit.edu}}

\abstract{The decay \Bsmm is very sensitive to contributions from new physics processes.
Thus the Tevatron and LHC experiments are hunting for an observation of a \Bsmm signal.
In this article the updated search for \Bsmm and \Bdmm by the CDF experiment is presented.
The CDF result was received with great interest because an excess over the background expectation is seen, although of modest statistical significance and still consistent with the prediction of a standard model signal and other experimental results.}

\FullConference{The 2011 Europhysics Conference on High Energy Physics-HEP 2011,\\
		July 21-27, 2011\\
		Grenoble, Rh\^one-Alpes France}

\begin{document}

The flavor-changing neutral current decays \Bsmm and \Bdmm are strongly suppressed in the standard model (SM) because they require higher order diagrams.
Their branching ratios are predicted to be $(3.2 \pm 0.2) \times 10^{-9}$ and $(1.0 \pm 0.1) \times 10^{-10}$, respectively \cite{Buras:2010mh}.
Physics beyond the SM could significantly alter these values, making the decays a sensitive probe for new physics.
So far the \Bmm decay was not observed in any experiment and the upper limits on the branching ratios provide stringent constraints on the parameter space of several new physics models.

Using a data sample of 7~fb$^{-1}$ collected at the Tevatron $p\bar{p}$ collider, the CDF experiment presented an updated search for \Bsmm and \Bdmm decays \cite{Aaltonen:2011fi}.
Compared to the previous CDF result \cite{Aaltonen:2007kv} based on a data sample of 2~fb$^{-1}$, the sensitivity is improved by the increased integrated luminosity, an extended muon trigger acceptance, and a more powerful signal-to-background discrimination.

Muons are detected by two types of chambers in the central (C) rapidity region of $|\eta|<0.6$ and in the forward (F) region of $0.6<|\eta|<1$.
Dimuon events are triggered by a pair of muons with either both muons in the central region (CC) or one in the central and one in the forward region (CF).
An improved understanding of the trigger performance allowed to increase the acceptance by 20\% compared to the previous analysis which excluded regions of rapidly changing efficiency.

The two muon tracks are fitted to a common vertex and baseline requirements~\cite{Aaltonen:2007kv} are applied.
%The selection uses muon identification information, the transverse momenta of the muon tracks and the \B candidate, the decay time significance, the pointing angle between the \B momentum and flight direction, and the isolation of the dimuon pair.
In the remaining sample, the dimuon invariant mass signal region from 5.169 to 5.469 GeV is blinded and the final selection is optimized using simulated signal events and background events from mass sidebands.
With a neural network~\cite{Feindt:2006pm} of 14 input variables the background could be suppressed about twice as efficiently as in the previous analysis for the same signal acceptance.
Overtraining was avoided by using statistically independent training and validation samples.
It was checked that the neural network selection does not bias the mass distribution and, using \BuJpsiK decays, that the distribution of the network output, $\nu_N$, is well described by the simulation.

To exploit the varying signal-to-background ratio as a function of $\nu_N$ and dimuon invariant mass, \Mmm, the data is split into bins in both variables.
An optimization on the expected upper limit yields eight bins in $\nu_N$ and five bins in \Mmm.
Expected background rates and signal efficiencies are calculated for each of the 40 bins for the CC and CF sample.

Since the combinatorial background has a smooth dependency on \Mmm, it is estimated by an extrapolation from the mass sidebands to the signal region using a linear function.
The sideband region is restricted to $\Mmm>5$ GeV to avoid contributions from $B \rightarrow \mu^+\mu^-X$ events.
Background from $B$ meson decays to two hadrons that are misidentified as muons could peak in the signal region.
Its contribution is estimated by taking the kinematic distribution from simulation, the normalization from known branching ratios, and momentum dependent misidentification probabilities measured in $D^0 \rightarrow K^-\pi^+$ decays.
The \Bhh background yield is highest in the most sensitive neural network bin.
In the \Bd search window of the CC sample it is comparable to the combinatorial background in this bin.
The \Bhh contribution in the \Bs search window is about an order of magnitude lower.
The background estimation procedure was verified on control samples consisting of muon pairs with like-sign change, with negative proper decay time, or with relaxed muon identification requirements.

With the background estimate at hand, one can determine whether the data is consistent with background only or whether there is an excess.
To obtain a quantitative result on the \Bmm branching ratio in the form of a limit or measurement, the signal yield is normalized to the number of \BuJpsiK decays which are kinematically similar.
The relative efficiency is determined using simulation and muon reconstruction efficiencies measured in data.
Based on these numbers 1.9 \Bs signal events are expected in the CC+CF sample in the SM.
This is the first time a non-negligible contribution from the SM signal is expected in a \Bsmm search at the Tevatron.

\begin{figure}
\centering
\includegraphics[width=0.85\textwidth]{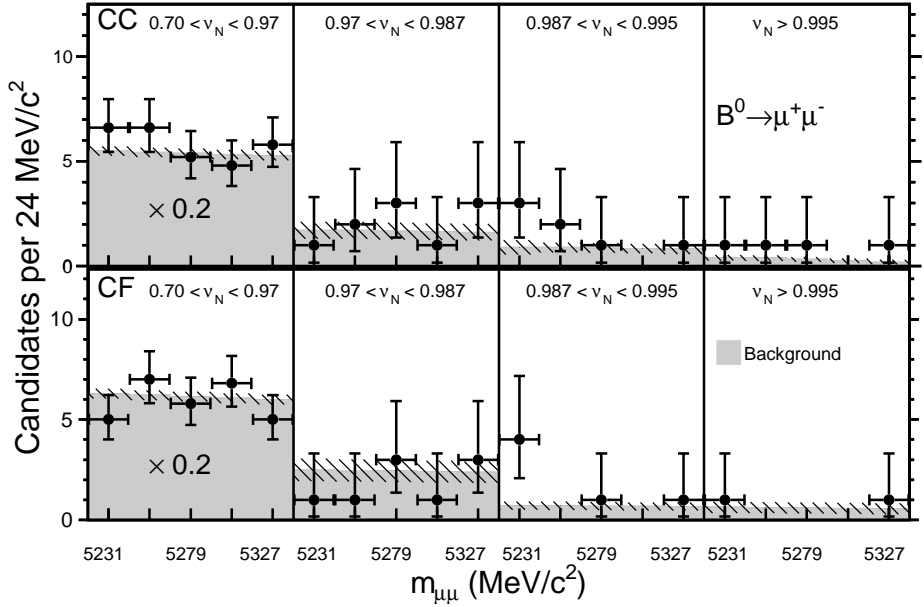}
\caption{Number of observed events in the \Bd search region versus \Mmm and $\nu_N$ for the CC (top) and CF (bottom) sample compared to the background estimation.
The systematic uncertainty on the background is indicated by the hatched area.}
\label{fig:BdResult}
\end{figure}

The unblinded data for the \Bd search window is shown in Fig.~\ref{fig:BdResult}.
It is well consistent with the background estimation and a 95\% C.L. limit of $\Br(\Bdmm)<6.0 \times 10^{-9}$ is obtained.

\begin{figure}
\centering
\includegraphics[width=0.85\textwidth]{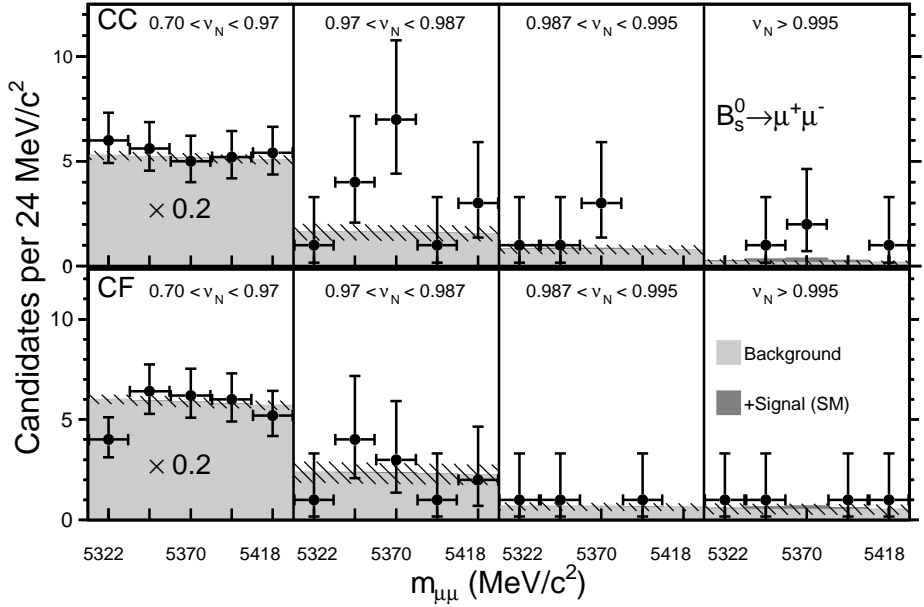}
\caption{Number of observed events in the \Bs search region versus \Mmm and $\nu_N$ for the CC (top) and CF (bottom) sample compared to the background+SM signal estimation.
The systematic uncertainty on the background is indicated by the hatched area.}
\label{fig:BsResult}
\end{figure}

In the highest $\nu_N$ bin of the \Bs search window the data shows an excess over the background estimate as can be seen in Fig.~\ref{fig:BsResult}.
The level of agreement is evaluated using a likelihood ratio whose expected distribution is determined with pseudoexperiments.
For the background only hypothesis a p-value of 0.27\% is calculated.
Taking into account the SM contribution, it increases to 1.9\%.

Since the third highest $\nu_N$ bin in the CC sample also shows an excess, the analysis was carefully checked for effects which could produce such a peak.
Peaking \Bhh background is unlikely because it is not seen in the \Bd search window.
A \Bsmm signal at this bin is also unlikely because the $\nu_N$ distribution is well modeled by the simulation for \BuJpsiK decays.
Peaking combinatorial background would have been discovered in the selection cross-checks or shown up in the \Bdmm search.
Thus a statistical fluctuation in the third highest $\nu_N$ bin is the most likely explanation.
Given the large number of 80 bins considered in this analysis such a deviation can be expected.
If only the two highest $\nu_N$ bins are used to evaluate the agreement with the background expectation, p-values of 0.66\% and 4.1\% are obtained for the background-only and background+SM hypotheses, respectively.

Given the marginal agreement with the background-only hypothesis, CDF performs a measurement of $\Br(\Bsmm)=1.8^{+1.1}_{-0.9} \times 10^{-8}$ and sets a 90\% C.L. interval of $4.6 \times 10^{-9} < \Br(\Bsmm) < 3.9 \times 10^{-8}$.
This result is consistent with the 90\% C.L. limits quoted by CMS~\cite{Chatrchyan:2011kr} and LHCb~\cite{BmumuLHCb}.
Although the excess is of modest statistical significance, it is likely that \Bsmm signal events contribute to the observed data.
In this sense it can be regarded a first indication of a \Bsmm signal which should not be confused with a claim for new physics.

\end{document}